\input harvmac

\def\p{\partial}
\def\ap{\alpha'}
\def\half{{1\over 2}}

\def \dag {\dagger}

\def \d {\delta}
\def \eps {\epsilon}

\def \lam {\lambda}

\def \s {\sigma}

\def \th {\theta}

\def \la {\langle}
\def \ra {\rangle}
\def \R {{\bf R}}

\def \Dc {{\cal D}}
\def \Psit {\tilde{\Psi}}

\Title{hep-th/0006024}{\vbox{\centerline{Black Holes and Spacetime Physics 
in String/M Theory}}}
\vskip20pt

\centerline{Miao Li}
\vskip 10pt
\centerline{\it Institute of Theoretical Physics}
\centerline{\it Academia Sinica}
\centerline{\it Beijing 100080} 
\centerline { and}
\centerline{\it Department of Physics}
\centerline{\it National Taiwan University}
\centerline{\it Taipei 106, Taiwan}
\centerline{\tt mli@phys.ntu.edu.tw}

\bigskip

In addition to briefly reviewing recent progress in studying black hole 
physics in string/M theory, we describe several robust features pertaining 
to spacetime physics that one can glean by studying quantum physics of 
black holes. In particular, we review 't Hooft's S-matrix ansatz
which results in a noncommutative horizon. A recent construction
of fuzzy $AdS_2$ is emphasized, this is a nice toy model for fuzzy
black hole horizon. We demonstrate that this model captures some
nonperturbative features of quantum gravity.

\Date{May 2000}

\nref\bl{S. Chandrasekhar, The Mathematical Theory of Black Holes,
Clarendon Press, Oxford University Press; S. W. Hawking and 
G. F. R. Ellis, The Large Scale Structure of Space-time, Cambridge:
Cambridge University Press, 1973.}
\nref\bek{J. D. Bekenstein, Nuovo Cim. Lett. 4 (1972) 737; Phys. Rev. D7 (1973)
2333; D9 (1974) 3292.}
\nref\haw{S. W. Hawking, Comm. Math. Phys. 43 (1975 199; J. B. Hartle and 
S. W. Hawking, Phys. Rev. D13 (1976) 2188.}
\nref\inf{S. W. Hawking, Phys. Rev. D14 (1976) 2460; Comm. Math. Phys. 87 
(1982) 395.}
\nref\bsp{T. Banks, L. Susskind and M. E. Peskin, Nucl. Phys. B244 (1984) 125.}
\nref\transp{'t Hooft, Nucl. Phys. B256 (1985) 727; Int. J. Mod. Phys. A11 
(1996) 4623.}
\nref\suss{L. Susskind, L. Thorlacius and J. Uglum, hep-th/9306069, 
Phys. Rev. D48
(1993) 3743; L. Susskind, hep-th/9308139, Phys. Rev. D49 (1994) 6606.}
\nref\holo{G. 't Hooft, gr-qc/9310026; L. Susskind, hep-th/9409089, J. Math. 
Phys. 36 (1995) 6377.}
\nref\mald{J. Maldacena, hep-th/9711200, Adv. Theor. Math. Phys. 2 (1998) 231.}
\nref\sv{A. Strominger and C. Vafa, hep-th/9601029, Phys. Lett. B379 (1996) 
99.}
\nref\joe{J. Polchinski,  hep-th/9510017, Phys. Rev. Lett. 75 (1995) 4724.}
\nref\mbh{T. Banks, W. Fischler, I. R. Klebanov and L. Susskind, 
hep-th/9709091,
Phys. Rev. Lett. 80 (1998) 226; hep-th/9711005, JHEP 9801 (1998) 008;
I. R. Klebanov and L. Susskind, hep-th/9709108,
Phys. Lett. B416 (1998) 62.}
\nref\hml{G. Horowitz and E. Martinec,  hep-th/9710217, Phys. Rev. D57 (1998) 
4935;
M. Li, hep-th/9710226, JHEP 9801 (1998) 009; M. Li and E. Martinec, 
hep-th/9801070.} 
\nref\giant{J. McGreevy, L. Susskind and N. Toumbas, hep-th/0003075; M. Li,
hep-th/0003173.}
\nref\fuzzy{P. M. Ho and M. Li, hep-th/0004072; P. M. Ho and M. Li, 
hep-th/0005268.}
\nref\rvbl{J. Maldacena, hep-th/9607235; A. Peet, hep-th/9712253, Class. Quant.
Grav. 15 (1998) 3291; K. Skenderis, hep-th/9901050.}
\nref\hs{G. Horowitz and A. Strominger, Nucl. Phys. B360 (1991) 197.}
\nref\cm{C. Callan and J. Maldacena, hep-th/9602043, Nucl. Phys.
B472 (1996) 591.}
\nref\mls{J. Maldacena and L. Susskind, hep-th/9604042, Nucl. Phys. B475 (1996)
679.}
\nref\dm{S. Das and S. Mathur, hep-th/9606185 Nucl. Phys. B478 
(1996) 561; A. Dhar, G. Mandal and S. Wadia,
hep-th/9605234, Phys. Lett. B388 (1996) 51.}
\nref\ja{J. Maldacena and A. Strominger, hep-th/9609026, Phys. Rev. D55 (1997)
861.}
\nref\fdbl{J. Maldacena and A. Strominger, hep-th/9603060,
Phys. Rev. Lett. 77 (1996) 428; C. V. Johnson, R. R. Khuri and
R. C. Myers, hep-th/9603061, Phys. Lett. B378 (1996) 78.}
\nref\spbl{J. C. Breckenridge et al., hep-th/9602065, Phys. Lett.
B391 (1997) 93; J. C. Breckenridge et al., hep-th/9603078, Phys. Lett.
B381 (1996) 423.}
\nref\as{P. C. Aichelburg and R. U. Sexl, Gen. Rel. Grav. 2 (1971) 303;
W. B. Bonner, Comm. Math. Phys. 13 (1969) 163.}
\nref\sst{N. Seiberg, L. Susskind and N. Toumbas, hep-th/0005015.}
\nref\ty{T. Yoneya, p. 419 
in ``Wandering in the Fields", eds. K. Kawarabayashi and 
A. Ukawa (World Scientific, 1987) ;
see also p. 23 in ``Quantum String Theory", eds. N. Kawamoto
and T. Kugo (Springer, 1988).}
\nref\tamiaki{T. Yoneya, Mod. Phys. Lett. {\bf A4}, 1587(1989).}
\nref\ly{M. Li and T. Yoneya, hep-th/9611072, Phys. Rev. Lett. 78 (1997)
1219.}
\nref\lyrev{M. Li and T. Yoneya, ``Short-distance Space-time Structure
and Black Holes in String Theory: A Short Review of the Present
Status, hep-th/9806240, Jour. Chaos, Solitons and Fractals (1999)}
\nref\tamy{T. Yoneya, hep-th/0004074.}
\nref\lsuss{L. Susskind, Phys. Rev.  D49 (1994) 6606.}
\nref\gross{D. Gross and P. Mende, Nucl. Phys.  B303, 407(1988).}
\nref\dkps{ M. Douglas, D. Kabat, P. Pouliot and S. Shenker,
Nucl. Phys. B485(1997)85.}
\nref\bfss{ T. Banks, W. Fischler, S. H. Shenker and L. Susskind,
Phys. Rev. D55(1997)5112.}
\nref\almy{H. Awata, M. Li, D. Minic and T. Yoneya, hep-th/9906248;
C. S. Xiong, hep-th/0003292.}
\nref\ms{J. Maldacena and A. Strominger, hep-th/9603069, Phys. Rev.
Lett. 77 (1996) 428.}
\nref\madore{J. Madore,
``Introduction to Noncommutative Differential Geometry
and Its Physical Applications'',
Cambridge U. Press, 2nd (1999).}
\nref\kt{I. R. Klebanov and A. A. Tseytlin, hep-th/9604166.}
\nref\thh{G. 't Hooft, ``Quantization of Point Particles
in $2+1$ Dimensional Gravity and Space-Time Discreteness'',
gr-qc/9601014.}
\nref\yang{C. N. Yang, ``On Quantized Space-Time'',
Phys. Rev. 72 (1947) 874.}
\nref\aads{A. Strominger, hep-th/9809027, JHEP 9901 (1999) 007.}
\nref\bg{T. Banks, M. Green, hep-th/9804170,
JHEP 9805 (1998) 002.}

\newsec{Introduction}

Quantum mechanics and Einstein's general relativity are both regarded as 
achievements of highest form in physics in twentieth century. Both have 
profoundly reformed
our view of physical world, as well as have had many applications in explaining
numerous observational facts. Yet, despite many heroic efforts, it has proven
formidable feat to achieve to put both theories in a single theoretical 
framework. String theory has long been held by many as the most promising 
candidate for such a framework. Recent progress in unifying different string 
models into a single theory, the M theory, and in using string theory to 
understand certain aspects
of quantum black holes, has reinforced this optimistic belief. 
String/M theory can not claim that its goal of unifying quantum mechanics 
and gravity has been achieved,
mainly due to the lack of a fundamental formulation capable of describing all
different situations even in principle.

To formulate a theory unifying quantum mechanics and general relativity,
one feels that in the end the new theory must incorporate fundamental
features of both theories. Quantum mechanics is a general kinematic 
theory supposed to be valid no matter what forces are involved, or what the
detailed dynamics is. On the other hand, general relativity was a new
invention of Einstein to reformulate gravity theory. It is dynamic by 
nature. However this is a very special dynamics, since it deeply involves
spacetime which is taken as a fundamental ingredient in any kinetic
theory. Thus one can not simply apply quantum mechanics to general 
relativity by treating the latter as just another dynamic system. 
In other words, the new theory must be kinematic as well as dynamic at the 
same time. It is very likely that when one attempts to include
other forces or interactions into this theory, these interactions
cannot be arbitrary, and must be an inseparable part of the holistic
theory. This is surely the spirit of Einstein when he pursued
a unified theory and dreamed to derive quantum from geometry in his
late life. 
String theory seems to have this quality: When strings
or things such as D0-brane partons are quantized, one gets 
fluctuations of spacetime for free along with other matter particles.

Many questions concerning the fate of spacetime become much sharpened 
in black holes
\bl. It was observed by Bekenstein that a Schwarzschild black hole seems to
represent the maximal entropy state of all physical systems that can be 
contained in a spherical
region, and this entropy is proportional to the area of its horizon \bek. 
A black hole is not really black at the quantum level \haw, it emits Hawking 
radiation. It
turns out that in the entropy formula both the Planck constant and the Newton 
constant
play a role, indicating that a proper understanding of the origin of the 
entropy requires a theory of quantum gravity. This necessity is enhanced 
by the riddle of information loss \inf. It was believed by some that in 
the process of formation and
evaporation of a black hole, quantum gravity effects play a minor role, so that
the pure quantum state will evolve into a mixed state, and unitarity is lost. 
This is at odds with very general considerations of energy conservation \bsp, 
and it seems hard to build a theory with violation of unitarity at the 
Planck scale only.
Among others, 't Hooft and Susskind have argued that infalling particles 
as well
as Hawking particles of finite energy are extremely boosted to a stationary 
observer near the horizon \refs{\transp, \suss}, so that strong gravitational 
interaction among
trans-Planckian particles is essentially involved. Based on this and the assumption
that unitarity is preserved all the time, 't Hooft postulated his S-matrix anstatz \transp,
and noncommutativity of spacetime first appeared. Susskind advocated that some
basic properties of string theory seem to fit the unconventional physics near a 
black hole. In particular, 't Hooft's reasoning leads to a drastic truncation of the
Hilbert space, since for instance two high energy particles with opposite momenta
within a fixed impact parameter is indistinguishable from a small black hole. This
is one of reasons to propose the holographic principle \holo. This principle
has gained much popularity recently through the celebrated Maldacena conjecture
\mald.

String theory has been successful in explaining the origin of entropy for various
extremal and near extremal black holes, starting with the work \sv. This 
lends much support to the belief that string theory not only overcomes 
the perturbative divergences but also encodes correctly some nonperturbative 
features of spacetime
physics. All constructions of extremal black holes involve D-brane 
configurations 
\joe. The D-brane theory is always weakly coupled when a reliable calculation can
be done. In such a situation, spacetime curvature is large and therefore gravitional
interactions are strong. Nevertheless one can still trust Bekenstein-Hawking formula
since it is protected by supersymmetry. In some occasions, one still can trust D-brane
theory even in its strong coupling limit, due to some nonrenormalization theorems.
The black hole becomes macroscopic thus classic geometry emerges. Maldacena conjecture
may be regarded, in somewhat technical view, as embodiment of infinitely many
nonrenormalization theorems.

How to treat a Schwarzschild black hole in string/M theory still remains a big open problem,
despite much work done in the context of matrix theory \refs{\mbh, \hml}.
Physics can in principle be studied in matrix theory, however geometry as well
as other related physical quantities are difficult to be recovered, just as in the D-brane
approach to extremal black holes. It appears that new insights are needed in order
to build a more fundamental and transparent picture. We suspect that these insights
will have much to do with quantum geometries which already showed up in 
't Hooft's work, and with further understanding of holographic principle. 
Some recent proposals on fuzzy spheres and fuzzy anti-de Sitter spaces will 
provide a good starting point for pursuit in this direction 
\refs{\giant, \fuzzy}. We have every 
reason to believe that much progress will be made in near future.

In the next section, we start with a brief account of thermodynamic 
properties of a Schwarzschild black hole and an extremal black hole, 
with an emphasis on the unusual
physics viewed by a stationary observer. Sect.3 reviews a calculation
of extremal black holes in string theory. We will be brief again, 
since there exist several reviews on this subject \rvbl. We will discuss 
some work of 't Hooft in sect.4 with an emphasis on the quantum geometry 
arising from the S-matrix ansatz. We will also show in sect.5 that 
spacetime uncertainty in string theory has
a similarity to the noncommutativity of 't Hooft. Schwarzschild black holes
in matrix theory will be discussed in sect.6, here we will  see that geometry
is hard to study, this perhaps is the generic problem in all
holographic theories. Finally we present some  recent progress in
understanding physics in fuzzy $AdS_2$ in sect.7. 

We hope that the present short
review article will help to call more attention to several very interesting
subjects presented here, in addition to serving as a concise guide to
more detailed original works.

Most of time we will use the natural unit in which $c=\hbar=1$.
The Newton constant $G$ is not set to one, to emphasize
one of the important length scales when the interactions are
switched on. Without warning we sometimes reinstall $\hbar$
to show the quantum origin of some effects.

\newsec{Thermodynamics and Other Properties of Black Holes}

Given suitable initial boundary conditions, it can be proven that formation of
a black hole with event horizon is inevitable. For ordinary matter with 
interactions (or equation of state) dictated by known forces, it is well-known
that a black hole forms of a burnt-out star of a few solar masses. It is possible
that much smaller black holes can form of nonordinary matter such as scalar
excitations in an early universe. It is also possible to form a microscopic 
black hole in a violent collision of extremely energetic particles.

The most studied, yet the most mysterious black hole from string theory viewpoint
is the Schwarzschild black hole of mass $M$. Its metric
\eqn\met{ds^2=-(1-({r_0\over r})^{D-3})dt^2+(1-({r_0\over r})^{D-3})^{-1}dr^2
+r^2d\Omega^2_{D-2}}
solves the D dimensional Einstein equations in vacuum, where $d\Omega^2_{D-2}$ is
the metric on a unit round $D-2$ dimensional sphere. At $r=r_0$, $g_{00}$ 
degenerates, thus the red-shift factor becomes infinity viewed by an 
outside observer sitting at $r>r_0$. We will focus on the case $D=4$ in this
section. $r_0=2GM$ in this case, and we always set $c=1$. Define
\eqn\tt{\rho =r+r_0\ln ({r\over r_0}-1)}
for $r>r_0$, the metric reads
\eqn\confc{ds^2=-(1-{r_0\over r})dX^+dX^-+r^2d\Omega_2^2,}
where $X^\pm=t\pm\rho$.  The first part of the metric vanishes at the horizon.

The Hawking temperature is given by
\eqn\ht{T_H={1\over 4\pi}\p_rg_{00} (r=r_0)={1\over 4\pi r_0},}
here we have set both the Boltzmann constant $k=1$ as well as the Planck 
constant $\hbar =1$. Thus the temperature always has an energy unit.
It is easy to see from \ht\ that the Hawking temperature is proportional
to $\hbar$, so the origin of Hawking temperature is quantum mechanical.
If the gravitational barrier is negligible for Hawking radiation,
we see that the typical wave-length of Hawking radiation is the order
of $r_0$. This fact alone already indicates that a black hole is
a unusual statistical system. Note that the Hawking temperature
is the one viewed by a distant observer. The local temperature for
an observer located at $r$ is easily obtained by the red-shift formula
\eqn\loct{T(r)=T_Hg_{00}(r)^{-1/2}.}
For $r$ close to $r_0$, we have $T(r)=1/(2\pi d(r))$, where
$d(r)\sim2\sqrt{r_0(r-r_0)}$ is the proper distance from the horizon.
These formulas becomes more and more accurate when $r_0$ gets larger
and larger. In fact, $T=1/(2\pi d)$ is the formula for the Unruh 
temperature in the Rindler space
\eqn\rind{ds^2=-e^{y^++y^-}dy^+dy^-,}
where $\exp y^\pm =X^\pm$ and $X^\pm$ are\ the flat space light-cone
coordinates. The proper distance at $\rho=\half (y^++y^-)$ from
the horizon is $e^\rho$ and the Unruh temperature is $1/(2\pi)
e^{-\rho}$. The fact that the geometry of a large black hole is close
to the Rindler space makes it clear that study of quantum physics
near a black hole will also bring about insights for physics in
the flat spacetime.

The entropy of the black hole can be deduced from the first law
of thermodynamics, assuming that indeed the black hole can be treated
as a thermodynamic system. In order to see the quantum nature
of the entropy, we reintroduce $\hbar$ in $T$. Substitute $E=M
=r_0/(2G)$ and $T=\hbar/(4\pi r_0)$ into $dE=TdS$, we find
$dS=dA/(4G\hbar)$ where $A=4\pi r_0^2$ is the area of the horizon.
Integrating this formula we have
\eqn\ent{ S={A\over 4G\hbar}={A\over 4l_p^2},}
we ignored an additive constant term which is presumably microscopic
and nonuniversal. Note that $S$ diverges in the limit $\hbar
\rightarrow 0$, agreeing with the fact that a black hole is
black in the classical limit. We deduced the entropy using the
first law of thermodynamics, since it is the shortest derivation
we can imagine. The original argument leading to \ent\ does not
require thermodynamics at all. In fact, Bekenstein was motivated
by the analogy of growth of the total horizon area in many dynamic
processes to identify $A$ with $S$ \bek. The second law of 
thermodynamics is always valid.

The appearance of the Planck length $l_p$ in the entropy formula
suggests strongly to many people that the origin of this entropy
must lie in quantum gravity, since $l_p$ is the length scale at
which quantum gravity effects become visible. The innocent looking
of our derivation of \ent\ may tempt people to suspect that
the area formula may be derived within quantum field theory in
a fixed curved background, since Hawking radiation itself was
derived in this way. The fact that this expectation is wrong
can be seen from the following simple argument. Consider for
instance a massless particle outside the horizon. By the usual
statistical mechanics, the local entropy density is $aT^3(r)$.
The integral
\eqn\divi{a\int T^3(r)g_{00}^{-1/2}r^2dr=a T_H^3\int r^2
g_{00}^{-2}dr}
diverges linearly near $r=r_0$. It also diverges for large $r$.
The latter divergence can be viewed from contribution of
the bulk matter surrounding the black hole, and can be separated
from the contribution near the horizon by a more careful analysis. 
The former divergence is a genuine
UV property. It can be formally cut-off by introducing a cut-off
on the proper distance $d(r)\sim l_p$. Thus one would obtain
an entropy formula similar to \ent. This formal cut-off calls
for a quantum gravity interpretation, and is termed sometimes
as the brick-wall model.

A physical cut-off would typically violate Lorentz invariance,
so the brick-wall model can hardly be taken literally. One may
take the necessity of introducing a cut-off as implying that
beyond the Planck temperature, the number of degrees of freedom
becomes smaller and smaller. An alternative explanation seems
even more attractive: Spacetime is fundamentally noncommutative,
so when one probes the high energy regime, one reaches back to
the long distance physics, thus, an integral such \divi\ is not
well defined when both $1/T$ and $d(r)$ become small. It must be
replaced by something else. We will see that indeed there is a
Lorentz invariant quantum geometry which potentially offers
such a formula.

Another puzzle concerns the origin of the Hawking radiation. The
life time of the black hole can be obtained by integrating
$dM/dt\sim -T^4_HA\sim -1/r^2$. We have $t\sim r_0^3/l_p^2$. 
Suppose the Hawking radiation originate from a place a proper
distance $d$ away from the horizon. The time for a S-wave to
propagate out is given by
\eqn\timep{t=\int g_{00}^{-1}dr \sim r_0\ln {r_0\over r-r_0}
\sim 2r_0\ln {r_0\over d}.}
If the brick wall is really there, it forms a perfect reflecting
mirror, so for the majority of radiation to be able to escape
it must be placed at $d$ such that the above time is comparable to
th life time of the black hole. We deduce from this 
\eqn\toos{ d\sim r_0\exp(-{r_0^2\over 2l_p^2}),}
this is an absurdly small distance compared to the Planck length
for a macroscopic black hole.

An extremal or a near extremal black hole has been of central interest
in the past few years in string theory. The simplest one comes from 
the Reissner-Nordstr\"om black hole. The 4D metric is
\eqn\cbl{ds^2=-f(r)dt^2+f^{-1}(r)dr^2+r^2d\Omega_2^2}
with
\eqn\fbl{f(r)=1-{2GM\over r}+{Q^2\over r^2}={1\over r^2}(r-r_+)(r-r_-),}
where $r_\pm =GM\pm\sqrt{(GM)^2-Q^2}$ are the radii of the outer and 
inner horizons respectively. $Q$ is the electric charged carried by
the black hole. The inequality $GM\ge |Q|$ must hold to
avoid naked singularity. When the equality holds, the black hole
is extremal. The time and radial coordinates exchange their roles when one crosses
the outer horizon. Their roles are exchanged one more time when
the inner horizon is crossed.

The Hawking temperature is obtained using formula \ht, it
is 
\eqn\hht{T_H={1\over 4\pi r_+^2} (r_+-r_-).}
It becomes zero for the extremal black hole. Thus it appears that
the nonextremal charged black hole continues Hawking radiation
until it rests as an extremal black hole. To obtain the entropy of
a charged black hole, we use
\eqn\nfirt{dM= T_HdS +VdQ,}
where $V$ is the static potential at the outer horizon and assumes the
value $V=Q/r_+$. With this formula and the Hawking temperature \hht,
\nfirt\ can be seen to be integrable and to yield
\eqn\cent{S={\pi r^2_+\over l_p^2}={A\over 4\l_p^2}.}
Again the area law holds.

It appears that an extremal black hole is somewhat strange in that
it has a large degeneracy, despite the fact that its Hawking temperature
vanishes. Also, the brick-wall model seems to work again. But when one
holds fixed the proper distance the brick wall from the outer horizon,
the coordinate distance $r_*-r_+$ goes to zero as $1-r_-/r_+$. 
An anti-de Sitter space is obtained from the extremal black hole
by taking a scaling limit, see sect.7. 
Thus, the charged black hole is valuable for testing any idea
implementing Planck scale physics.

\newsec{Extremal Black Holes in String Theory}

String theory distinguishes itself from other approaches to quantum 
gravity by postulating the existence of closed as well as open strings. 
The discovery of dualities makes it clear that for consistency other
extended objects must be included too. The most important such objects
are D-branes \joe. These branes carry charges with respect to 
Ramond-Ramond antisymmetric tensor fields. Depending on which dual
description one uses, strings can also be regarded as D-branes. 
D-branes are more fundamental in that their description contains
open string excitations. These excitations are important in the counting
of the number of states for various extremal black holes.

The single most important fact making string theory different in constructing
black hole solutions is the existence of the dilaton field. This
is a massless scalar field when the ground state has enough supersymmetry.
Its expectation value determines the strength of string interactions.
There are Schwarzschild black hole solutions in string theory. These
are the same as in Einstein theory since in a vacuum the dilaton is a
constant. However, various abelian gauge fields are coupled to the
dilaton in one way or another, charged black holes are rather different
from, say, the standard Reissner-Nordstr\"om black hole \hs. Thus,
to obtain an extremal black hole with a macroscopic horizon, it is
necessary to switch on several different charges, and the minimal number
of charges is 3. The first such black hole is obtained in 5 dimensions
\sv. Four charged black holes can be constructed in 4 dimensions.

We discuss only the 3 charged black hole in 5 dimensions, following \cm.
Starting with IIB 
theory and compactifying it on $T^5$, there is
a abelian gauge field $C^{(2)}_{a\mu}$ resulting from $C^{(2)}$,
the rank 2 R-R field.
A wrapped D-string along $X^a$ carries its electric charge. 
Another abelian gauge field, $C^{(6)}_{1,\dots,5,\mu}$ results
from the dual of $C^{(2)}$. A D5-brane wrapped around $T^5$ carries
its electric charge. Two unbroken supersymmetry conditions
$\epsilon=\gamma^{0a}\tilde{\epsilon}$ and $\epsilon=\gamma^{01\dots
5}\tilde{\epsilon}$ are compatible if $a$ is one of $1,\dots, 5$.
That is, the bound state of $N_5$ D5-branes wrapped around 
$T^5$ and $N_1$ D-strings wrapped around a circle of $T^5$ is a 
BPS state. The residual SUSY is $1/4$ of the number of original
supersymmetry. Take $a=1$.

We need one more charge to construct a black hole. This is achieved
by adding momentum modes along $X^1$, namely along the D-string
direction. This introduces a further constraint on unbroken SUSY 
$\epsilon=\gamma^{01}\epsilon$, $\tilde{\epsilon}=\gamma^{01}
\tilde{\epsilon}$. This means that both $\epsilon$ and $\tilde{
\epsilon}$ are positive eigenstate of $\gamma^{01}$. Combined
with the D-string constraint, $\epsilon=\tilde{\epsilon}$.
Thus there are 8 unbroken super-charges. Finally the D5-brane
constraint eliminates half of them. The BPS black hole preserves
$4$ super-charges.

To see that this is a black hole, we need the metric:
\eqn\tend{\eqalign{ds^2&=(H_1H_5)^{-1/2}(-dt^2+dX_1^2+(H_p-1)(dt-dX_1)^2)+
H_1^{1/2}H_5^{-1/2}(dX_2^2+\dots +dX_5^2)\cr
&+(H_1H_5)^{1/2}(dr^2+
r^2d\Omega_3^2),\cr
e^{2\phi}&=g^2H_1H_5^{-1},}}
where $H_i$ are harmonic functions in 5 dimensions, 
$H_i=1+r_i^2/r^2$, where the parameter $r_i^2$ is proportional
to the corresponding charge. And the R-R fields
\eqn\rrf{C^{(2)}_{01}={1\over 2}(H_1^{-1}-1),\quad
F_{ijk}={1\over 2}\epsilon_{ijkl}\p_l H_5,}
where $i,j,k,l$ are indices tangent to the 4 open spatial dimensions. 
Let $(2\pi)^4V$ denote the volume of $T^4$ orthogonal to the D-strings,
and $R_1$ the radius of $X^1$. It is easy to see that
\eqn\param{r_1^2={gN_1\over V},\quad r_5^2=gN_5,\quad 
r_p^2={g^2N_p\over R_1^2V},}
where the momentum along $X^1$ is $N_p/R_1$. We have set $2\pi\ap=1$.
For fixed $V$ and $R_1$,
we see that all sizes $r_i$ become macroscopically large when
$gN_1\gg 1$, $gN_5\gg 1$, and $g^2N_p\gg 1$. We call this region of
the parameters the black hole phase.

When reduced to 5D, the Einstein metric reads
\eqn\fived{ds^2=-(H_1H_5H_p)^{-2/3}dt^2+(H_1H_5H_p)^{1/3}\left(
dr^2+r^2d\Omega_3^2\right).}
From the component $G_{00}$ we see that $r=0$ is the horizon, since
the red-shift factor becomes infinity at this point.
The Bekenstein entropy is easy to calculate, either by using the 8 
dimensional horizon if the hole is treated as living in 10 dimensions,
or by using the 3 dimensional horizon when it is treated as living in 5 
dimensions. It is relatively simpler to use the 5D metric. The horizon
area is given by $2\pi^2[r^2(H_1H_5H_p)^{1/3}]^{3/2}$ when the
limit $r\rightarrow 0$ is taken.  Thus $A_3=2\pi^2r_1r_5r_p$,
and $r=0$ is not a point. The 5D Newton constant is $G_5=
g^2/(4VR_1)$, so the entropy is
\eqn\entr{S={A_3\over 4G_5}=2\pi\sqrt{N_1N_5N_p},}
a nice formula. 

It can be shown that all 5 dimensional black holes preserving $1/8$
of supersymmetry can be rotated into above black hole using U-duality,
here the U-duality group is $E_6$. If one can count the entropy
microscopically for one of them, then others must have a microscopic
origin too on the count of U-duality. For instance, a 5D black hole
in IIA theory is obtained by performing T-duality along $X^1$. The
hole is built with D4-branes, D0-branes bound to them, and string
winding modes around the dual of $X^1$. Now this has a simple M 
theory interpretation, the D4-branes get interpreted as fivebranes
wrapped around the M circle, winding strings get interpreted as
membranes wrapped around the M circle, and D0-branes are M momentum
modes. Thus, the hole is built using fivebranes intersecting
membranes along a circle with momentum modes running along this
circle.

Come back to the IIB 5D black hole. The simplest account of the
microscopic picture goes as follows. The D-strings are bound to 
D5-branes, and they live on the Higgs branch in the weak string
coupling limit, thus can oscillates only in the 4 directions along
D5-branes. If the size of $V$ is much smaller than $R_1$, the
oscillation is effectively described by a 1+1 conformal field theory.
The fluctuations correspond to wiggling of the open strings
stretched between D5-branes and D-strings, thus there are $4N_1N_5$
such bosons. Due to supersymmetry, there are also the same number of
fermions. The theory is therefore a conformal field theory with
central charge $6N_1N_5$. Since in a CFT a fluctuation is either
left-moving or right-moving, and we restrict our attention to BPS
states, there are only right-moving modes which contribute
to the total momentum $N_p/R_1$. Thus, $N_p$ is the oscillator
number. We are therefore interested in the coefficient of
$q^{N_p}$ in the expansion of the following partition function
\eqn\partiti{Z=\left(\prod_{n=1}{1+q^n\over 1-q^n}\right)^{4N_1N_5},}
and it is given, after a saddle point calculation, by 
$\exp(2\pi\sqrt{N_1N_2N_p})$, that is, the entropy agrees exactly
with \entr.

There is subtlety involved in the above calculation, making it
invalid for large $N_1$ and $N_5$. A cure of this problem is
provided by the fractionation mechanism, whose details we will not 
run into here \mls.

The Hawking temperature of the 5D extremal black hole is zero, so
it does not Hawking radiate. To obtain a nonextremal black hole,
we need to add some anti-charges to the hole. The metric \tend\
is modified by a nonextremal factor $f(r)=1-r_0^2/r^2$ to
the part $dt^2$, and $f^{-1}$ to the part $dr^2$. The location of the 
horizon is shifted to $r_0$. The simplest way to obtain this
metric is by adding momentum modes moving in the opposite direction
along $X^1$ compared to the existing modes. Now the Hawking
temperature is proportional to the square root of the number of the
added modes. Assuming that the Hawking radiation comes mainly
from combination of the left-moving modes and the right-moving
modes, a careful calculation by Das and Mathur \dm\ shows that
the perturbative string calculation reproduces exactly the correct
Hawking's black-body formula. It is even more striking that
a further perturbative string calculation reproduces the correct
grey-body factor \ja, which is just the energy-dependent absorption
cross section.

All the above discussions can be extended to the 4D extremal and 
near extremal black holes \fdbl, and to charged, spinning black
holes \spbl.

Now matter how nice the results one can obtain using the D-brane
technology, one can not help but feel that there is something
crucial missing. One does not have a geometric picture at all,
since the D-brane account is more or less holographic. With
the advent of Maldacena conjecture, understanding emergence
of geometry has become more urgent. In this regard, it is perhaps 
useful to go back where one started in first place.

\newsec{'t Hooft's S-matrix Ansatz}

Having seen impressive progress made in string theory explaining some
of the quantum properties of near extremal black holes, we come
back to issues concerning the Schwarzschild black holes. The main
reason for doing this is not only that these black holes are still
poorly understood in string theory, but also that understanding these
black holes will bring in new insights and perhaps it is about the 
right time to look back at some points having great potential in
near future.

One interesting approach is 't Hooft's S-matrix ansatz \transp. 
't Hooft postulates that quantum mechanics is always valid during
the whole process of formation and evaporation of a black hole, and
that the trans-Planckian regime can not be ignored when one
considers the effects of incoming particles on the Hawking radiation.
As we already pointed out in sect.2, very close to the horizon, due
the the enormous red-shift factor, an infalling particle as well
as an outgoing particle with finite energy measured at infinity
are boosted to very high energies. Whether they are massive or
massless is immaterial. To see the mutual effect between an incoming
particle and an outgoing particle, let us examine the gravitional
field produced by a shock-wave in the flat spacetime first.

Consider an massless particle moving in the direction $x^3$. For
a left-moving particle, its wave function is a function of $x^+
=t+x^3$. The stress tensor thus has a nonvanishing component
$T_{++}$. A shock-wave is defined by the characteristic that the
energy is concentrated at a definite value, that is, $T_{++}
\sim \delta (x^+)$, here for convenience we choose the concentrating
point to be $x^+=0$. Physically, the massless particle passes
the point $x^3=0$ at $t=0$. Since spacetime is 4 dimensional,
we also need to specify the location of the particle in the
transverse space $(x^1,x^2)$. Again for simplicity we take the
location to be the origin. Due to the rotational invariance
in the transverse space, it is easy to convince oneself that
the metric produced by the shock-wave assumes the form
\eqn\swm{ds^2=-dx^+dx^-+\delta(x^+)h(r)(dx^+)^2 +dx^idx^i,}
where $x^i$ runs through $x^1, x^2$, and $r^2=(x^1)^2+(x^2)^2$.
The metric is flat away from $x^+=0$.
For $T_{++}=\delta (x^+)T(x^i)$, the Einstein equation
reduces to
\eqn\eine{\Delta h(x^i)=8\pi G T(x^i).}
Consider the special case $T(x^i)\sim delta^2(x)$. Since
by the definition
$$p_+=\int T_{++}dx^+d^2x,$$
we have $T_{++}=p_+\delta (x^+)\delta^2(x)$. And the solution
to the Einstein equation is 
\eqn\asm{h=4Gp_+\ln r.}
\asm\ together with \swm\ is called Aichelburg-Sexl metric \as.

The effect of the metric \swm\ on a right-moving particle can
be seen easily for a massless particle. The classical trajectory
in a flat spacetime is specified by $dx^-=0$. However, in
a background with a nonvanishing component $g_{++}$, the trajectory
is changed to $dx^-=g_{++}dx^+$. For $g_{++}$ of the form
\swm, the trajectory is $x^-=const$ on the both sides separated
by $x^+=0$. There is a jump when the right-moving particle
crosses $x^+=0$, the jump is
\eqn\jump{\Delta x^-= 4l_p^2p_+\ln r,}
if the location of the right-moving particle on the transverse space
is $r$, we have also replaced $G$ by $l_p^2$. 

Apparently, the result \jump\ is a purely classical result, 
despite the way we write it. In fact $l_p^2p_+$ can be roughly
viewed as the measure of the gravitional size, since it is
roughly the Schwarzschild black hole radius if there is a 
right-moving particle with the same energy colliding with the
left-moving one and forming a black hole. The validity
of \jump\ requires $r$ to be larger than a certain scale.
Lacking a detailed quantum gravity theory, we can only guess
that this scale can be either $l_p$ or $l_p^2p_+$ itself.

We generalize the above analysis to the geometry near a black hole
horizon. We introduced tortoise coordinate \tt\ in sect.2,
although the interesting part of the metric \confc\ is
conformal to the flat metric, it is inconvenient to work
with $\rho$ since the horizon is located at $\rho=-\infty$. 
The more convenient coordinate system is Kruskal coordinates
$x^\pm$, the relations between the two systems are
simply
\eqn\rlt{x^\pm =\exp (\pm {X^\pm\over 2r_0}).}
The past horizon corresponds to $X^+=-\infty$, or $x^+=0$,
the future horizon corresponds to $X^-=\infty$ or $x^-=0$. 
An incoming particle moves with fixed $x^+$ and with
decreasing $x^-$, and an outgoing particle moves with
a fixed $x^-$ and with increasing $x^+$. The Schwarzschild
metric is 
\eqn\krus{ds^2={4r_0^3\over r}e^{-r/r_0} dx^+dx^- +r^2d\Omega_2^2.}
Denote $2A$ the prefactor in the front of $dx^+dx^-$ in
the above formula, the metric produced by an incoming
shock-wave takes the general form
\eqn\anst{ds^2=2Adx^+dx^--2\delta(x^+)AF(dx^+)^2 +r^2d\Omega_2^2,}
where $F$ is a function of $x^-$ and angular variables
of the sphere $S^2$. 't Hooft calculated the Ricci tensor
of the above metric, and found that the only nonvanishing
component is $R_{++}$. Let $\Delta$ be the Laplacian on
the unit sphere, Einstein equation reads simply
\eqn\einq{\Delta F-F=8\pi {G\over r_0^2} p_+\delta (\Omega),}
where $\delta (\Omega)$ is the delta function on the
unit sphere. 

If the incoming particle locates at the north pole of the
two sphere, the solution to \einq\ is
\eqn\solu{F(\theta)={4G\over r_0^2}p_+f(\theta)=-{4G\over
r_0^2}p_+ \sum_l 
{l+\half\over l(l+1)+1} P_l(\cos\theta ).}
The effect of of the incoming shock wave on an
outgoing particle is therefore a shift in $x^-$
\eqn\shim{\Delta x^-={4G\over r_0^2}p_+ f(\theta).}
This shift is a constant. However, in terms of $X^-$,
the shift is proportional to $\exp (X^-/(2r_0))$. For
an outgoing particle originating sufficiently close
to the future horizon, where $X^-=\infty$, this shift
is enormous.

Now we can construct a piece of the S-matrix in the background
of a black hole. A particle or a state of multiple-particles
coming from infinity near the horizon ($r-r_0\sim r_0$)
is described by $S_{in}$,
this part is governed more or less by the known theory. Similarly,
$S_{out}$ describes how outgoing particles leave from
the region $r-r_0\sim r_0$. A nontrivial piece, denoted by
$S_{hor}$, describes the effect of incoming particles on the
outgoing particle very close to the horizon. Thus the S-matrix
in a fixed black hole background is $S=S_{out}S_{hor}
S_{in}$. This splitting is approximate only. We are mostly
interested in $S_{hor}$. 

For outgoing particles, the phase is shifted by an amount
$\exp (-ip_-\delta x^-)$. Thus
\eqn\mout{\psi_{out}=e^{-i\int d\Omega P_-(\Omega)
\delta x^-(\Omega)}\psi_{in},}
where $P_-(\Omega)$ is the outgoing momentum density operator.
Using \shim, the above is rewritten
\eqn\fout{\psi_{out}=e^{-i\int d\Omega d\Omega' P_-(\Omega)
F(\Omega, \Omega')P_+(\Omega')}\psi_{in},}
where $P_+(\Omega')$ is the incoming momentum density operator.
The near hole S-matrix reads simply
\eqn\smt{S_{hor}={\cal N}\exp \left(-i\int d\Omega d\Omega' 
P_-(\Omega)F(\Omega, \Omega')P_+(\Omega')\right).}
This is the main result of 't Hooft.
The above S-matrix is not satisfactory, since it still assumes that
the transverse part of the horizon $S^2$ is a continuous surface,
and the function $F$ suffers a logarithmic divergence when two points
on $S^2$ get close. This can be cured by generalizing the shock-wave
of carrying only longitudinal momentum to one also carrying some
transverse momentum. One is lead to some kind of fuzzy sphere.
This is quite similar to the recently discovered fuzzy spheres in
the AdS/CFT correspondence \fuzzy.

The S-matrix ansatz elevates the classical result \shim\ to a 
quantum mechanic one. One may go one step further to
claim that \shim\ implies a commutator between $x^+$ and 
$x^-$:
\eqn\fhor{[x^+(\Omega), x^-(\Omega')]={4G\over r_0^2}f(\Omega, \Omega')
[x^+,p_+]={4iG\hbar\over r_0^2}f(\Omega, \Omega'),}
thus the Planck length $l_p^2=G\hbar$ automatically appears
after a simple incorporation of quantum mechanics.
We will see that the fuzzy $AdS_2$ model exhibits a very similar
commutation relation between the light-cone coordinates,
although the initial motivation for proposing this model
is quite independent of the shock-wave argument \fuzzy.

A physically nontrivial consequence of the S-matrix \smt\ is teleology.
An observer who sees an outgoing particle would deduce that an
incoming particle is affected: If one trace back along the trajectory
of the outgoing particle, it also induces an enormous shift on an
incoming particle. Thus an infalling observer would conclude
that what he sees is correlated with what a distant observer sees.
This leads to the conclusion that operators with spacelike separation
do not commute. A recent discussion on this phenomenon in a 
noncommutative field theory and string theory can be found in 
\sst.

\newsec{Spacetime Uncertainty in String/M Theory}

The first string revolution brought about powerful
perturbative tools in studying string theory, and
a classification of perturbative string vacua.
Although a number of interesting things concerning
the nature of space were discovered, such as
T-duality and mirror symmetry, the nature of space and
time was largely obscure. The only exception
was the pioneering work of Yoneya \ty, \tamiaki\
on a spacetime uncertainty relation. This
proposal had been ignored by the community
until the second string revolution. The first revival
of interest in this relation was a check in D-brane
dynamics \ly. This relation is rather stringy,
and its M theory generalization was proposed in
\lyrev. Further elaborations on this subject
are presented in a beautiful review \tamy.

In the perturbative formulation of string theory,
conformal invariance on the world-sheet plays
a fundamental role. It is therefore important
to extract from conformal invariance some
physical property which may survive interactions.
Here we follow the original approach of
\ty\ to derive a spacetime uncertainty relation.
Consider a parallelogram on the string world-sheet  and the
Polyakov amplitude for the mapping from the
world-sheet to a region of the target  space-time.
Let the lengths of the two orthogonal sides in the
world-sheet parallelogram be
$a, b$ in the conformal gauge
where $\dot{x} \cdot x'=0, \,
\dot{x}^2 +x'^2 =0$  and the
corresponding physical
space-time length be $A, B$, respectively.
Then apart from the
power behaved pre-factor, the amplitude
is proportional to
\eqn\abamp{
\exp [-{1\over \ell_s^2} ({A^2\over \Gamma}
+{B^2 \over \Gamma^*})]}
where
\eqn\recip{
\Gamma \equiv {a\over b}, \, \, \Gamma^* \equiv {b\over a} ,
\quad (\Gamma \Gamma^* =1).}
Due to the conformal invariance, the amplitude
depends on the Riemann sheet parameters only through
the ratio $\Gamma$ or $\Gamma^*$,
which are called the {\it extremal length} and the conjugate
extremal length, respectively.  Clearly, the relation
$\Gamma \Gamma^*=1$ leads to the uncertainty relation 
\eqn\stunc{
\Delta T \Delta X
 \sim \ell_s^2 \sim \alpha'}
taking the $A$ direction to be time-like
$\Delta T\sim \langle  A \rangle$ and
hence the $B$ direction to be space-like
$\Delta X \sim  \langle  B\rangle$.
The obvious relation $\Gamma \Gamma^*=1$ is the
origin of the familiar modular invariance of the
torus amplitudes in string theory.  The extremal
length is the most fundamental moduli
parameter characterizing conformal invariants,
in general.  Since arbitrary
amplitudes can be constructed by
pasting together many parallelograms, any string
amplitudes satisfy the above reciprocal relation
qualitatively.  Although this form looks
too simple as the characterization of the
conformal invariance, it has a virtue
that its validity  is very general, as we
will explain shortly,  and does not use the
concepts which depend intrinsically  on
perturbation theory.  Our proposal  is to
use this relation as one of possible guiding
principles towards nonperturbative
reformulation of string theory and M-theory.

The uncertainty relation \stunc\
is consistent with an
elementary property of strings that the energy of
a string is roughly proportional to its space-time length.
\eqn\hunc{
\Delta E \sim  {\hbar \over \alpha'}\Delta X_l .}
with $X_l$ being the length of a string measured
along its longitudinal direction.
Then the ordinary time-energy uncertainty relation
$\Delta T \Delta E \ge \hbar$ leads to \stunc.
It is important here to discriminate the length scales in the longitudinal
and transverse directions with respect to the string.
  As is well known, 
transverse length scale grows  logarithmically with energy used to probe
the strings.
This explains the linearly rising Regge trajectory
for the Regge-pole behavior in  high-energy peripheral
scattering.
The dual role of the time and the longitudinal
spatial lengths
is a natural space-time expression of the original
$s$-$t$ duality.
The particle exchange (Regge exchange
in the old language)  and
the resonance behaviors correspond to the
regimes, $\Delta X_l \rightarrow \infty$ and
$\Delta T \rightarrow \infty$, respectively.
Furthermore, the Regge behavior is
consistent with the existence of graviton,
since scattering amplitudes in general
are  expected to be roughly proportional to $\Delta X_l \sim \ell_s^2 
/\Delta T
\propto E$ which
implies, by adopting the argument
in \lsuss,  that the intercept $\alpha(0)$ of the leading Regge 
trajectory
is 2, from the relation $E \sim E^{\alpha(t)-1} $.

On the other hand,
in the high energy fixed-angle scatterings with
large $s$-and $t$-channel momenta studied
in detail in \gross,
we are trying to probe the region where both the time and the
spatial scales are small. Clearly, such a region is incompatible
with the space-time uncertainty relation. The exponential fall-off
of the perturbative string amplitudes in this limit
may be interpreted as a manifestation of this property.
According to the space-time uncertainty relation, at least
one of the two length scales, $\Delta X$ or
$\Delta T$,  must be larger than the
string scale $\ell_s$.  Therefore there is no degrees of
freedom left in this regime. However, when
one really tries to extract spacetime information
from the fixed angle amplitudes, one
finds that they are compatible with the relation
\stunc, as recently discussed in \tamy.
It is well known that any consistent theory of quantum gravity
must indicate lessening of the degrees of freedom
near the Planck scale where the quantum nature of gravity
becomes important.
We have seen that the space-time uncertainty relation
can indeed be a natural mechanism for this.
Qualitatively, it is also consistent with the
known high-temperature behavior of the
perturbative string amplitude, since the
high-temperature limit is effectively
equivalent to considering the limit $\Delta T\rightarrow 0$.

To check that the spacetime uncertainty relation
hold in D-brane dynamics, we consider the simplest,
perhaps most important case of D0-branes.
Consider the scattering of two D0-branes of mass $1/g_s\ell_s$ with
the impact parameter of order $\Delta X$ and the relative
velocity $v$ which is assumed to be much smaller than the
light velocity. Then the characteristic interaction time $\Delta T$ is
of order ${\Delta X\over v}$. Since the impact parameter is of the
same order as the longitudinal length of the open strings
mediating the interaction of the D-particles, we can
use the space-time uncertainty relation in the form
$$\Delta T \Delta X \sim \ell_s^2 \Rightarrow
{(\Delta X)^2 \over v} \sim \ell_s^2$$
This gives the order of the magnitude for the  minimum possible
distances probed by the D-particle scatterings with
velocity $v \, (\ll 1)$.
\eqn\veldis{
\Delta X\sim
\sqrt{v}\ell_s.}
To probe short spatial
distances, we have to use D-particles with small
velocity.  However, the slower the
velocity is and hence the longer  the interaction time is,
the larger is the spreading of the wave packet.
\eqn\dxdt{
\Delta X_w \sim
\Delta T\Delta_w v \sim
{g_s\over v}\ell_s,}
since the ordinary time-energy uncertainty relation
says that the uncertainly of the velocity is of
order $\Delta_w v\sim g_sv^{-1/2}$ for the time
interval of order $\Delta T \sim v^{-1/2}\ell_s$.
Combining these two conditions, we see that the
shortest spatial length is given by
\eqn\pleng{
\Delta X \sim  g_s^{1/3}\ell_s}
and the associated time scale is
\eqn\adt{
\Delta T \sim  g_s^{-1/3}\ell_s .}
\pleng\  is of course the
11 dimensional Planck length which
is the characteristic length of M-theory which was
first derived in the super YM context in \dkps.
As argued in \ly, it is actually possible to
probe shorter lengths than the Planck length
if we consider a D-particle in the
presence of many (=$N$) coincident D4-branes.

One can extract the M theory uncertainty
relation from the stringy one valid for D0-branes.
This derivation assumes the validity of matrix theory
\bfss. Note that a process involving individual D-partons necessarily
smears over the longitudinal direction, thus the uncertainty in
this direction $\Delta X_L=R$ is maximal. Relation \stunc\ is
rewritten as
\eqn\munc{
\Delta X_T\Delta X_L\Delta T\ge l_p^3,}
this relation  refers only to the fundamental
length scale in M theory,
the Planck length, thus it is a natural candidate for the generalized
uncertainty relation in M theory. We now argue
that relation \munc\  is the correct relation for a process
involving  a boosted cluster. It is trivially true for threshold
bound state, since it is just a boosted parton and according to
Lorentz invariance $\Delta X_L$ contracts, while $\Delta T$
is dilated by a same factor. An object carrying the same
amount of longitudinal momentum can be regarded as an excited state
of the threshold bound state, therefore intuitively as a probe
it can not probe a transverse distance shorter than a threshold
bound state can do. Thus, relation \munc\  must also hold
for such a probe. Note also that this relation is Lorentz invariant.

When the new relation \munc\ was proposed in
\lyrev, it was also checked that this relation is
valid in the AdS/CFT correspondence \mald, where
conformal invariance on the M-branes is essentially
employed. More recently, it was checked that this
relation is compatible with the stringy exclusion
principle, based on a remarkable dipole
mechanism proposed in \giant. We hope that a 
precise mathematical framework properly accommodating
\munc\ will offer a clue to a covariant formulation
of matrix theory. Some attempts to constructing new brackets
which may be relevant to the cubic uncertainty relation
can be found in \almy.

It remains to connect the more microscopic relations
\stunc, \munc\ to the noncommutative black hole 
horizon we discussed in the previous section.

\newsec{Schwarzschild Black Holes in Matrix Theory}

Matrix theory promises us a nonperturbative definition of M theory
in 11 dimensions, as well as toroidal compactification
on a torus $T^d$ with $d\le 5$. If so, a Schwarzschild
black hole must be in principle describable in matrix
theory. Indeed many semi-quantitative results were
obtained in this framework, such as the scaling law
between the mass of the black hole and its radius 
\mbh, \hml.
Nevertheless matrix theory has the reputation of unwieldy,
so by far it is impossible even to extract the standard
Schwarzschild geometry.

The simplest situation is when the radius of the black
hole, after boosted, matches the infrared
cut-off size in the longitudinal direction in matrix theory.
Naively, the radius of a black hole, like everything
else in special relativity, contracts in the longitudinal
direction with a boost: $r_0\rightarrow e^{-\beta}r_0$,
where $\beta$ is the boost parameter. For the
matrix theory to be effective, $r_0\gg R$.
The longitudinal momentum scales inversely
with $e^{-\beta}$, so $P_-\sim r_0M/R$. Since
in matrix theory, all longitudinal momentum is
carried by the partons, $P_-=N/R$, $N$ is the number of
partons, we have $N\sim r_0M$. This number
is the same magnitude of the black hole entropy,
since in $D$ dimensional spacetime, there are relations
\eqn\ddbl{S\sim {r_0^{D-2}\over G}, \quad M\sim {r_0^{D-3}
\over G},}
where $G$ is the D dimensional Newton constant. 
That the number of partons required matches
roughly the entropy of the black hole strongly
suggests that the partons are responsible for
counting of microscopic states. Also note that
for matrix theory to be able to accommodate the
black hole, $N$ is related to the minimal boost,
thus $N$ may be understood as the minimal
number of partons required to describe the
black hole microscopically.

It was later pointed out in \hml\ that the Lorentz
contraction really occurs to the size of
the black hole as seen by a distant observer, and
that the actual size of the horizon is not
changed, as suggested the purely geometric definition of the 
black hole horizon, which doesn't depend on which
coordinates system one uses. Interpreted by
a distant observer, who actually uses matrix theory
to describe the black hole, the size of the black hole
becomes larger for a probe near the horizon, and
this enlargement is due to the pressure exerted
by partons carrying longitudinal momentum.
Before running into any details, we already
see that it is going to be hard to study geometry
in matrix theory, since as an input, the
geometry is always fixed at spatial infinity,
and in order to study geometry generated
by sources, we need first of all define new
geometric quantities in matrix theory.

There is a very simple derivation of the relation
between the radius and the mass in matrix theory.
As we said above, the longitudinal radius is
not easy to see, however, the transverse radius
is not distorted by boost, thus can be 
seen directly in matrix theory. When $D=11$, consider the
effective interaction between two partons in
the leading order
\eqn\leadi{V=c{l_p^9\over R^3}{(v_1-v_2)^4\over r^7},}
where $c$ is a numerical constant. The interaction
is attractive, thus $c$ is negative, if $V$ is taken
as a contribution to the effective two-body
Hamiltonian. 

Here are two crucial points in reconstructing the black
hole data. First, every parton has a thermal
wave length comparable to the size of the hole.
This is a highly nontrivial assumption, as we know
that this is far from being generic in a thermal system.
Second, the hole may be regarded as a metastable
system such that one can treat it as a stable bound
state for all practical purposes. From the first
assumption, one gets
\eqn\vel{v\sim {R\over r_0},}
this is equal to the boost factor. From the second assumption,
one has the virial theorem
\eqn\vith{\half {v^2\over R}\sim N {l_p^9 v^4\over R^3r^7_0}.}
Combining these two equations, we derive
\eqn\nest{N\sim {r_0^9\over l_p^9},}
that is, the number of partons is approximately the
entropy of the black hole. This estimate is independent
of the previous ``fit the box" argument, thus we expect
that by combining that argument with the above
formula, we will get the relation between $M$ and 
$N$.  Alternatively, by substituting \nest\ into the
matrix Hamiltonian, one gets
\eqn\lce{H=E_{LC}\sim \half {N\over R}v^2\sim {Rr_0^7\over l_p^9},}
and the mass
\eqn\blm{M=\sqrt{P_-E_{LC}}\sim {r_0^8\over l_p^9}.}
This is precisely the mass formula for the hole.
This derivation is independent of the boost argument
we gave in the beginning of this section.
It is difficult to justify the boost argument 
in the present context, since Lorentz boost
properties are hard to study.

One way to probe the geometry near the lump
of partons as a black hole is to study probes. The simplest
probe is the D0-brane parton itself. This project
was initiated in the third paper of \hml. In a boosted
black hole geometry, the D0-brane action is
the generalized Born-Infeld action. In order to
probe the full geometry, one needs to calculate
the interaction between the probe and the lump up
to all orders in the double expansion in the velocity
and the distance. This is hard to do. Thus so far
it is fair to say that even the classical geometry of
the matrix black hole has not been extracted.

The even more interesting problem is to study the
quantum process of scattering a D0-brane parton
against the lump, in order to extract the quantum
geometry similar to the one proposed by 't Hooft.

\newsec{Fuzzy $AdS_2\times S^2$}

\subsec{Fuzzy $AdS_2\times S^2$ from AdS/CFT Correspondence}

Consider the near horizon limit of a 4 dimensional charged
black hole in string theory \ms.
For instance, by wrapping two sets of membranes and
two sets of M5-branes in $T^7$, one obtains a 4D charged,
extremal black hole \kt.
The brane configuration is as follows.
Denote the coordinates of $T^7$ by $x_i$, $i=1, \dots, 7$.
A set of membranes are wrapped on $(x_1, x_2)$,
another set are wrapped on $(x_3, x_4)$.
A set of M5-branes are wrapped on $(x_1, x_3, x_5, x_6, x_7)$,
the second set are wrapped on $(x_2, x_4, x_5, x_6, x_7)$.
By setting all charges to be $N$, one finds the metric of
$AdS_2\times S^2$ for $(x_0,x_8,x_9,x_{10})$:
\eqn\adsm{\eqalign{&ds^2=l_p^2\left(-{r^2\over N^2}dt^2+{N^2\over r^2}dr^2
+N^2d\Omega_2\right),\cr
&F=-N d\Omega_{1+1}
-N d\Omega_2,}}
where $l_p$ is the 4 dimensional Planck length,
$d\Omega_{1+1}$ and $d\Omega_2$ are the volume forms
on $AdS_2$ and $S^2$, respectively.
The field $F$ is just the linear combination of all
anti-symmetric tensor fields involved.
Note that here for simplicity,
we consider the most symmetric case in which
all the charges appearing in the harmonics $1+Q_il_p/r$
are just $N$ which in turn is equal to the number
of corresponding branes used to generate this potential.
As a consequence, the tension of the branes compensates 
the volume of the complementary torus.
This means that the size of each circle of $T^7$
is at the scale of the M theory Planck length.

The same space $AdS_2\times S^2$ can also be obtained by
taking the near horizon limit of the 4 dimensional
extremal Reissner-Nordtstr\"om solution.

In \fuzzy\  we proposed that the $S^2$ part of
the $AdS_2\times S^2$ space is a fuzzy $S^2$ \madore\
defined by
\eqn\stwo{
[Y^i,Y^j]=i\epsilon^{ijk}Y^k,}
where $Y_i$'s are the Cartesian coordinates of $S^2$.
(We use the unit system in which $l_p=1$.)
This algebra respects the classical $SO(3)$ invariance.

The commutation relations \stwo\
are the same as the $SU(2)$ Lie algebra.
For the $(2N+1)$ dimensional irreducible representation of $SU(2)$,
the spectrum of $Y_i$ is $\{-N, -(N-1), \cdots, (N-1), N\}$.
and its second Casimir is
\eqn\casm{
\sum_{i=1}^3 (Y_i)^2=N(N+1).}
Since the radius of the $S^2$ is $N l_p$
(in the leading power of $N$),
we should realize the $Y_i$'s as $N\times N$ matrices
on this irreducible representation of $SU(2)$.

One evidence for this proposal is the following.
For a fractional membrane wrapped on $(x_1,x_3)$ or $(x_2,x_4)$,
it is charged under the $F$ field generated by a set of M5-branes.
Denote the polar and azimuthal angles of $S^2$ by $(\th,\phi)$.
The stable trajectories of the membrane
with the angular momentum $M$ are discrete.
\eqn\fract{
\cos\th={M\over N},}
and they all have the same energy of $1/N$.
It follows that, since $M$ is conjugate to $\phi$,
$\cos\th$ and $\phi$ do not commute with each other
in the quantized theory.
The resulting Poisson structure on $S^2$ is precisely
that of the fuzzy sphere.

In \thh, it was proposed that in $2+1$ dimensions
the spacetime coordinates are quantized according to
\eqn\xyL{
[x,y]={i\over\cos^2\mu}L,}
where $\cos\mu$ is related to the mass of the particle,
and $L$ is the angular momentum on the 2 dimensional space.
To complete the algebra we write down the usual relations
\eqn\rinv{
[L,x]=iy, \quad [L,y]=-ix.}
This algebra is rotational invariant,
and its 3+1 dimensional generalization
was given by Yang \yang\ much earlier.
Note that this algebra \xyL\
was proposed based on general grounds for
a gravitational theory in 2+1 dimensions,
and we are content with the fact that
it is actually a consequence of
the algebra of the fuzzy $S^2$ \stwo\
for massless particles ($\cos\mu=1$),
where $Y_3$ acts on $Y_1$ and $Y_2$ as
the angular momentum.

In \fuzzy\ we further proposed that the $AdS_2$ part
is also quantized.
Let $X^{-1}, X^0, X^1$ be the Cartesian coordinates of $AdS_2$.
The algebra of fuzzy $AdS_2$ is
\eqn\sucomm{\eqalign{
&[X^{-1}, X^0]=-i X^1, \cr
&[X^0, X^1]=i X^{-1}, \cr
&[X^1, X^{-1}]=i X^0, }}
which is obtained from $S^2$ by a ``Wick rotation''
of the time directions $X^0,X^{-1}$.
The ``radius'' of $AdS_2$ is $R=N l_p$, so that
\eqn\xxr{
\eta_{ij}X^i X^j=(X^{-1})^2+(X^0)^2-(X^1)^2=R^2,}
where $\eta=$diag$(1,1,-1)$.
The isometry group $SL(2,\R)$ of the classical $AdS_2$
is a symmetry of this algebra,
and thus is also the isometry group of the fuzzy $AdS_2$.

For later use, we define the raising and lowering operators
\eqn\gne{
X_{\pm}\equiv X^{-1}\pm i X^0,}
which satisfy
\eqn\gcomm{
[X^1, X_{\pm}]=\pm X_{\pm}, \quad [X_+, X_-]=-2X^1,}
according to \sucomm.

The radial coordinate $r$ and the boundary time coordinate $t$
are defined in terms of the $X$'s as
\eqn\pcoor{
r=X^{-1}+X^1, \quad t={R\over 2}(r^{-1}X^0+X^0 r^{-1}),}
where we symmetrized the products of $r^{-1}$ and $X^0$
so that $t$ is a Hermitian operator.
The metric in terms of these
coordinates assumes the form \adsm.
It follows that the commutation relation for $r$ and $t$ is
\eqn\rt{
[r,t]=-iRl_p.}

The following simple heuristic argument also
suggests this commutation relation.
Consider a closed string in $AdS_2$.
(Since the space is one dimensional,
the closed string actually looks like an open string
with twice the tension.)
Take the Nambu-Goto action for a fractional string of tension $1/N$
and take the static gauge $t=p_0\tau$.
It follows that the action is
\eqn\inp{\eqalign{
S&={1\over 2\pi N\ap}\int_{-\infty}^{\infty}dt\int_0^{2\pi}d\sigma
    \sqrt{(p_0\dot{r}^2)} \cr
 &={1\over \pi N\ap}\int dt\int_0^{\pi} p_0|\dot{r}| \cr
 &={1\over N\ap}\int d\tau \dot{r}{\p t\over\p\tau},}}
where we have assumed that $\dot{r}>0$ for $0<\s<\pi$
and $\dot{r}<0$ for $\pi<\s<2\pi$.
Now repeating an argument of sect.5, we conclude that there
is the uncertainty relation $\delta r\delta t> Rl_p$.

\subsec{Properties of Fuzzy $AdS_2$}

One can realize the algebra \sucomm\
which is the same as the Lie algebra of $SL(2,\R)$,
on a unitary irreducible representation of $SL(2,\R)$.
The question is which representation is the correct choice.
Since the range of $X_1$ goes from $-\infty$ to $\infty$ for $AdS_2$;
when $R>1/2$, the proper choice is the principal continuous series,
Since we have $R=N>1$ for our physical system,
we should consider the principal continuous series only.
A representation in this series is labeled by two parameters
$j=1/2+is$ and $\alpha$, where $s, \alpha$ are real numbers,
and $0\leq\alpha<1$.
The label $j$ determines the second Casimir as
\eqn\secc{
c_2=\eta_{ij}X^i X^j.}
It follows from \xxr\ and $R=N$ that one should take
\eqn\jN{
j=1/2+iN.}

We set $\alpha =0$ to
focus on the case in which the reflection symmetry,
$X^1\rightarrow -X^1$ is not broken.
We will denote this representation 
by $\Dc_N$ and focus on this case in the following.

Functions on the fuzzy $AdS_2$ are functions of the $X$'s.
They form representations of the isometry group $SL(2,\R)$.
The three generators of the isometry groups act on $X$ as
\eqn\rotg{
[L_{ij}, X^k]=i(\d_j^k X_i-\d_i^k X_j),}
where $X_i=\eta_{ij}X^j$.
A very interesting property of the algebra of fuzzy $AdS_2$ is that
the action of the generators $L_{ij}$ is
the same as the adjoint action of the $X_k$.
That is,
\eqn\redr{
[L_{ij}, f(X)]=\eps_{ijk}[X^k, f(X)]}
for an arbitrary function $f(X)$.
The operators $L_{ij}$ act on the functions as differential operators.

The integration over the fuzzy $AdS_2$ is just the trace
of the representation $\Dc_N$
\eqn\iiner{
\int f(X)\equiv c\Tr (f(X))=c\sum_{n\in Z}\la n|f(X)|n\ra,}
where $c$ is a real number.
This integration is invariant under $SL(2,\R)$ transformations.
In the large $N$ limit, $c_2\gg 1$,
the trace can be calculated and its comparison with
an ordinary integration on the classical $AdS_2$
with metric \adsm\ shows that
$c=N$ in the leading power of $N$.
The inner product of two functions, as well as the norm of a function
are defined by integration over the fuzzy $AdS_2$ in the usual way:
\eqn\emt{
\la f(X)|g(X)\ra=\int f^{\dag}(X)g(X),
\quad \|f(X)\|^2=\int f(X)^{\dag}f(X).}

In view of organizing the functions into representations of $SL(2,\R)$,
in order to describe the boundary CFT dual to the bulk theory
on fuzzy $AdS_2$ via holography,
we find all functions corresponding to the lowest
and highest weight states in the principal discrete series.
The information about the underlying fuzzy $AdS_2$,
i.e., the value of $N$,
is encoded in the precise expressions of these functions.

Denote the lowest weight state by $\Psi_{jj}$, or just $\Psi_j$.
The states of higher weights $\Psi_{jm}$ ($m>j$)
in the same irreducible representation
are obtained as $[X_+,[X_+,\cdots[X_+,\Psi_j]\cdots]]$,
where $X_+$ appears $(m-j)$ times.
With some calculation, we find the explicit expressions
for the functions $\Psi_j$ as
\eqn\psif{
\Psi_j=\left({1\over X^1(X^1-1)+c_2}X_+\right)^j.}
In the large $N$ limit, using $c_2=R^2$ and
the following coordinate transformation
\eqn\glc{
X^1=R\cot(u^+ - u^-),
\quad X_{\pm}={R\over\sin(u^+ - u^-)}e^{\mp i(u^+ + u^-)},}
one finds
\eqn\cpsif{
\Psi_j\rightarrow \left({e^{-iu^+}-e^{-iu^-}\over 2iR}\right)^j,}
where $u^+$, $u^-$ are the coordinates appearing in the global
parametrization of $AdS_2$,
in agreement with the classical result \aads.

Let
\eqn\Imdef{
I_{jm}\equiv \Tr (\Psi^{\dag}_{jm}\Psi_{jm}),}
then the normalized states are
\eqn\npsif{
\Psit_{jm}\equiv {1\over \sqrt{cI_{jm}}}\Psi_{jm}.}
As a normalized basis of an $SL(2,\R)$ representation,
they satisfy
\eqn\rlo{\eqalign{
{[X_+,\Psit_{jm}]}&=a_{j(m+1)}\Psit_{j(m+1)}, \cr
{[X_-,\Psit_{jm}]}&=a_{jm}\Psit_{j(m-1)},}}
where
\eqn\cnu{
a_{jm}=\sqrt{m(m-1)-j(j-1)}.}

For a field $\Phi$ in the bulk of the fuzzy $AdS_2$,
one can decompose it into the basis functions $\Psit_{jm}$ as
\eqn\jm{
\Phi(X)=\sum_{jm}\phi_{jm}\Psit_{jm}(X),}
where $\phi_{jm}$ are the creation/annihilation operators
of the physical state with the wave function $\Psit(X)$.
By holography, these states are identified with those
in the boundary theory, which is a one-dimensional theory.

An interesting question for a wave function on
noncommutative theory is how to define expectation values.
For instance, should the expectation value of $X_1$ for
a wave function $\Psi$ be
\eqn\clok{
(1)\quad \int \Psi^{\dag}X^1\Psi,
\quad (2)\quad\int \Psi X^1\Psi^{\dag},
\quad  (3)\quad
\half\int(X^1\Psi^{\dag}\Psi+\Psi\Psi^{\dag}X^1)?}
The answer is that it depends on how one measures it.
If one measures the $X^1$ location of the wave function
according to its interaction with another field $\Phi$
under control in the experiment,
and if the interaction is described in the action by a term like
\eqn\shn{
\int \Psi^{\dag}\Phi\Psi,}
then we expect that the choice (1) is correct.
But if the interaction is written differently,
the definition of expectation value should be modified accordingly.

\subsec{Interaction in Fuzzy $AdS_2$}

To see how the noncommutativity of the fuzzy $AdS_2$
incorporate physical data, presumably including the effect of
string quantization on $AdS_2$,
we consider an interaction term in the action of the bulk theory
of the form
\eqn\intv{
S_I=\lam\int\Phi^{\dag}_1\Phi_2\Phi_3,}
where $\lam$ is the coupling constant for this three point interaction.

Expanding all three $\Phi_i$'s as \jm\ in the action,
one obtains the vertex
\eqn\vertex{
c\lambda\Tr (\Psit^{\dag}_{j_1 m_1}\Psit_{j_2 m_2}\Psit_{j_3 m_3})}
for the three states $(\phi_1)_{j_1 m_1}$, $(\phi_2)_{j_2 m_2}$
and $(\phi_3)_{j_3 m_3}$.
Obviously, due to the isometry, the vertex vanishes unless
$m_1=m_2+m_3$.

For simplicity, consider the special case where all three
states participating the interaction are lowest weight states.
Then the vertex \vertex\ in question is $\lambda V_{m_1 m_2 m_3}$,
where (for $m_1=m_2+m_3$)
\eqn\Vtwo{
V^2_{m_1 m_2 m_3}={1\over c}{I_{m_1}\over I_{m_2}I_{m_3}}.}
After considerable calculations,  we find that
\eqn\ccoef{
I_m={(2m-2)!\over ((m-1)!)^2}
\left[\Pi_{k=1}^{m-1}{1\over j^2-1+4c_2}\right]I_1,}
where
\eqn\ione{
I_1={\pi\over\sqrt{c_2-1/4}}\tanh\left(\pi\sqrt{c_2-1/4}\right).}

We therefore have the large N expansion of the vertex \Vtwo.
Using \jN, one finds (for $m_1=m_2+m_3$)
\eqn\longe{
V^2_{m_1 m_2 m_3}={\cal N}_{m_2 m_3}
{\left[\Pi_{j_2=1}^{m_2-1}(1+j_2^2/4N^2)\right]
\left[\Pi_{j_3=1}^{m_3-1}(1+j_3^2/4N^2)\right]\over
8\pi^2 N^2\left[\Pi_{j_1=1}^{m_1-1}(1+j_1^2/4N^2)\right]}
\left(1+2\sum_{n=1}^{\infty}e^{-2\pi nN}\right),}
where
\eqn\mlong{
{\cal N}_{m_2 m_3}={[2(m_1-1)]! [(m_2-1)!]^2 [(m_3-1)!]^2
\over [(m_1-1)!]^2 [2(m_2-1)]! [2(m_3-1)]!}.}
With the possibility of corrections to \jN\ of order $(1/N)^0\sim 1$,
the $1/N$ expansion of the vertex is of the form
\eqn\Vm{
V_{m_1 m_2 m_3}\sim {K\over N}\left(1+
\sum_{n=1}^{\infty}{a_n\over N^{2n}}+
\sum_{n=1}^{\infty}e^{-2\pi nN}\sum_{k=0}^{\infty}
{b_{nk}\over N^{2k}}\right).}

This expression is reminiscent of a correlation function
in the case of type IIB strings on $AdS_5\times S^5$ \bg.
It calls for an analogous interpretation.

The $1/N^2$ expansion in \Vm\ suggests that
the coupling constant in $AdS_2$ is of order $1/N^2$.
This is indeed the case.
The 11 dimensional Newton constant is just $1$ in Planck units.
Compactifying on $S^2\times T^7$ of size $4\pi N^2$
results in a dimensionless effective Newton constant
of order $1/N^2$ in $AdS_2$.

The overall factor of $1/N$ in \Vm\
is what one expects for a three-point correlation function,
since for a large $N$ theory, the string coupling
constant is proportional to $1/N$, it appears in
the three-point coupling if the connected two-point
function is normalized to one.

Finally, we identify the terms in \Vm\ proportional to
$\exp(-2\pi nN)$ as contributions from instantons.
This implies that the action of a single instanton equals $2\pi N$.
We have just argued that the string coupling constant $g_s$
is of order $1/N$.
If the instanton action is $2\pi/g_s$,
it is precisely $2\pi N$ as we wish.

Similarly, we can consider $n$-point interaction vertex
in the bulk theory on $AdS_2$:
\eqn\npv{
S_n=\lam\int\Phi^{\dag}_1\Phi_2\cdots\Phi_n.}
The leading dependence of the vertex will be $1/N^{n-2}$,
which is exactly what it should be for an $n$-point function
in string theory with coupling constant $g_s\sim 1/N$.

We conjecture that for M theory compactified on $AdS_2\times S^2$,
the perturbative as well as nonperturbative effects
of string quantization are encoded
in the noncommutativity of the fuzzy $AdS_2\times S^2$,
in the sense that the low energy effective theory
is most economically written as a field theory on
this noncommutative space.

\subsec{A Shock-Wave Argument}

Although we already argued for the spacetime uncertainty
in the fuzzy $AdS_2$ from the general stringy uncertainty,
it should be interesting to compare the way 't Hooft
introduces spacetime noncommutativity in his S-matrix
ansatz. The method is similar to that in sect.4.

Use the global coordinates. The metric
induced by a left-moving shock-wave assumes the form
\eqn\shock{
ds^2=-e^{2\phi}du^+du^-+h(du^+)^2,}
where 
$$e^{2\phi}={4R^2\over \sin^2(u^+-u^-)}.$$
The scalar curvature is perturbed by a term
\eqn\pref{
{1\over 2}e^{-2\phi}\p_-(e^{-2\phi}\p_-h),}
and the Einstein equation with a constant negative 
curvature is solved provided
\eqn\tsolu{
h=\cot(u^+-u^-)g(u^+)+f(u^+).}
In the full 4 dimensions, we expect another Einstein
equation of the form $G_{++}=8\pi G_4T_{++}$, where
$G_4$ is the 4D Newton constant. Now $G_{++}\sim
h{\cal R}$, ${\cal R}$ is the scalar curvature. For a 
stress tensor $T_{++}$ proportional
to $\delta (u^+-x^+)$, the only solution is $g(u^+)=0$
and
\eqn\ssolu{
h\sim l_p^2\delta (u^+-x^+).}
The proportionality constant is determined by how the
stress tensor is normalized. For a S-wave shock-wave
smeared over $S^2$, it is  $p_+/R^2$. However,
the dipole mechanism of \fuzzy\ seems to indicate
that a shock-wave must be localized on a strip
on $S^2$ whose area is proportional to $Nl_p^2$. If true,
we expect
\eqn\nstt{
\int du^+ T_{++}\sim {p_+\over Rl_p},}
and this leads to
\eqn\fnst{
h\sim Rl_pp_+\delta (u^+-x^+).}

Now the shift on $u^-$ induced on a right-moving particle 
by the shock-wave is
\eqn\shsh{
\Delta u^-\sim {p_+\over N}\sin^2(x^+-u^-)}
as can be computed using \shock. This shift
suggests a commutator
\eqn\scom{
[u^+,u^-]\sim {i\over N}\sin^2(u^+-u^-),}
the one that is compatible with our fuzzy $AdS_2$ model.

\newsec{Conclusions}

We have amassed a few scattered aspects of the fuzzy spacetime
ranging from some elementary study of quantum horizon of
black holes by 't Hooft, to stringy spacetime uncertainty,
to fuzzy anti-de Sitter space. It can not be over-emphasized
that further and deepened study of all these aspects is one
of most urgent tasks in string/M theory. Here, instead of
pointing out problems already well-defined without much
a quandary, we ask a few questions.

\noindent 1. In quantum mechanics, observables are operators 
generally noncommuting. However, one rarely associates  
an observable to time. In this vein, we ask:
What is the most fundamental meaning of 
noncommutativity of space and time? It appears that
this notion challenges our usual understanding of both general
relativity and quantum mechanics. On the one hand, now 
an event is not a well-defined concept, thus challenging
the very foundation of general relativity.  On the other
hand, if time is not a usual number, one needs to 
extend quantum mechanics either in the Heisenberg form
or in the Schr\"odinger form, since in both one equates
the time derivative with the Hamiltonian operator.

\noindent 2. Clearly spacetime uncertainty is a generic feature
of string/M theory, do we need to formulate string/M theory
in a manifestly noncommutative fashion? 

\noindent 3. What is the relation between noncommutative horizon
of black holes and stringy noncommutativity? The latter sounds
more microscopic, while the former is more effective but presumably
includes nonperturbative information.

\noindent 4. Can one derive the entropy and other physical
quantities (such as modified Hawking radiation) of black holes
from noncommutativity?

\noindent 5. Does noncommutativity in string theory violate
causality by an arbitrary amount when the energy is increased,
presumably at a nonperturbative level?

\noindent 6. How to derive holography from noncommutativity?
Or does there exist a bulk formulation which incorporates
noncommutativity explicitly?

\noindent \dots

Acknowledgments. This work was supported by a grant of NSC and by a 
``Hundred People Project'' grant of Academia Sinica.  I thank
P. M. Ho and T. Yoneya for enjoyable collaborations on some subjects
presented here. This review article is prepared as a sketch of lectures
to be presented at the workshop on string theory
and noncommutative geometry at Center for Advanced Study of Tsing Hua 
University. I thank Y. S. Wu and Z. Xu for urging me to write
it up. We feel it is perhaps the right time to put this article to
the Archive and hope to have opportunity to expand it after the workshop,
and hopefully, after more developments occur in near future.

\vfill
\eject

\listrefs
\end